# Anomalously large measured thermoelectric power factor in $Sr_{1-x}La_xTiO_3$ thin-films due to $SrTiO_3$ substrate reduction

Matthew L. Scullin,[1,2] Choongho Yu,[1,3] Mark Huijben,[2,5] Subroto Mukerjee,[5] Jan Seidel,[5] Qian Zhan,[1] Joel Moore,[5] Arun Majumdar[1,4], R. Ramesh[2,5]

[1] Materials Sciences Division, Lawrence Berkeley National Laboratory, Berkeley, CA 94720

[2] Department of Materials Science and Engineering, University of California, Berkeley, Berkeley, CA 94720

[3] Department of Mechanical Engineering, Texas A&M University, College station, TX 77843

[4] Department of Mechanical Engineering, University of California, Berkeley, Berkeley, CA 94720

[5] Department of Physics, University of California, Berkeley, Berkeley, CA 94720

We report the observation that thermoelectric thin-films of La-doped $SrTiO_3$ grown on $SrTiO_3$ substrates yield anomalously high values of thermopower to give extraordinary values of power factor at 300K. Thin-films of $Sr_{0.98}La_{0.02}TiO_3$, grown via pulsed laser deposition at low temperature and low pressure (450°C, $10^{-7}$Torr), do not yield similarly high values when grown on other substrates. The thin-film growth induces oxygen reduction in the $SrTiO_3$ crystals, doping the substrate *n*-type. It is found that the backside resistance of the $SrTiO_3$ substrates is as low (~12Ω/□) as it is on the film-side after film growth.

Few material systems promise a compelling pathway to thermoelectric power factors significantly higher than that of bismuth telluride in bulk. While nanostructuring provides a means of increasing thermoelectric efficiency $zT$ through phonon engineering and reduction of lattice thermal conductivity,[1] selection of a next-generation material suitable for nanoscale improvement of thermoelectric efficiency may require a set of electronic transport properties such as that which is found in oxide semiconductors. As opposed to many narrow-gap semiconductor thermoelectrics whose thermopower $S$ are limited by increasing carrier concentrations so as to drive conductivity $\sigma$ higher, oxides—due to their carriers' sometimes extraordinarily large effective masses—are limited in $\sigma$ by mobility. Because power factor $S^2\sigma$ is more heavily weighted in thermopower, materials with large carrier effective mass and thus large thermopower provide perhaps the most interesting pathway to high $zT$. Strontium titanate, well-studied for its ferroelectricity[2], superconductivity[3], resistive switching[4], interfacial effects with other oxides[5,6,7,8], and low-dimensional thermoelectricity[9], can achieve very large effective mass (~4-16$m_e$) and thermopower (~0.8 mV/K) at 300 K[10]; it is also highly electronically tunable through a wide range of doping on either the A- or B-site or with oxygen vacancies. In addition, recent work on both the thermoelectric properties of confined layers of doped $SrTiO_3$ (STO) and interfaces of STO with other oxides[11] offer unique approaches to increasing $zT$ in these systems. In this letter we present findings that indicate an extremely high measured thermopower (>1 mV/K) is achieved at room temperature in Lanthanum-doped STO thin-films grown on STO substrates without any interfacial or confinement effects, due to a clear electronic contribution from the reduced substrate that must be accounted for even at low thin-film growth pressures for short times.

Based on both bulk data of La-doped $SrTiO_3$[12] and theory[13], we selected 2% La doping ($Sr_{0.98}La_{0.02}TiO_3$), the concentration which yields highest power factor in bulk. Films were grown from a dense polycrystalline, stoichiometric ceramic target onto $SrTiO_3$ (STO) (001), STO (001) / Si (110), $LaAlO_3$ (LAO) (001), $(LaAlO_3)_{0.3}$-$(Sr_2AlTaO_6)_{0.7}$ (LSAT) (001), and $NdGaO_3$ (NGO) (110) single-crystal substrates. Films were grown at a chamber base pressure of $1\times10^{-7}$ Torr and at 450°C at a pulsed laser energy density of 1.75 $J/cm^2$. Additionally, reflective high-energy electron diffraction (RHEED) oscillations were present throughout the entire growth of all SLTO films on STO, indicating 2D layer-by-layer film growth during the entire deposition process.

Both scanning transmission electron microscopy (STEM) and X-ray diffraction (XRD) data (Fig. 1) from the thin-films on STO indicate that they are single-phase, quasi-homoepitaxial, nearly single-crystal perovskite, and are also highly crystalline and single-phase on LSAT, STO/Si, LAO, and NGO (not shown in Fig. 1) substrates. Reciprocal space mapping of the 013 peak of the thin-films on STO reveal that the films are lattice-matched to the STO substrate along the *a-* and *b*-axes (but not necessarily on other substrates) at all thicknesses and distorted ~2.5% from the bulk out-of-plane along the *c*-axis on STO, and similarly distorted on all other substrates[14] (Fig. 1b). The full-width-at-half-maximum (FWHM) of the θ-2θ and ω (rocking-curve) 002 thin-film peaks in XRD show that films on STO substrates are of better crystal quality than those on LSAT, LAO, STO/Si, and NGO substrates, with crystallinity improving as the bulk lattice mismatch between the SLTO film and substrate shrinks to nearly zero in the case of STO (Fig. 1a). The in-plane lattice mismatch between the SLTO (001) ($a_{bulk\,(001)}$ = 3.9058 Å) films and the STO (001) ($a_{bulk\,(001)}$ = 3.9056 Å)[15],

LSAT (001) ($a_{bulk\,(001)}$ = 3.8688 Å)[16], NGO (110) ($a_{bulk\,110}$ = 3.8669 Å)[17], Si (110) ($d_{bulk\,(110)}$ = 3.84 Å), and LAO ($a_{bulk\,(001)}$ = 3.7911 Å)[18] films are 0.005, 0.94, 0.99, 1.68, and 2.93%, respectively, and the film FWHM in ω scale with them monotonically.

The ideal substrate for the growth of La doped-SrTiO$_3$, SrTiO$_3$ itself, becomes doped n-type to be electrically conductive when oxygen vacancies are created in the material[19]. Normally, an SrTiO$_3$ (001) single-crystal substrate subjected to the thermodynamic conditions of our thin-film growth for samples presented in this paper ($p_{O2}$ = 10$^{-7}$ Torr, 450°C, 10-60 minutes) will not become appreciably reduced nor conductive[20]; nevertheless only light doping of the substrate is required to give it a relatively small total resistance. The 500 μm-thick substrate (2.5x10$^3$ times the volume of the thin-films) requires a resistivity only 10$^{-3}$ times that of the film's resistivity to become a significant factor in any transport measurement assuming a uniform distribution of oxygen vacancies, and it must also be considered that short-circuit conductive paths with even lower resistivity may form along dislocations[21] or oxygen vacancy clusters[22]. To achieve a resistivity in the STO substrate (mobility $\mu_H$ ~ 5 cm$^2$V$^{-1}$s$^{-1}$)[23] of 2 Ω-cm at 300K—roughly 10$^3$ times greater than the bulk value of the Sr$_{0.98}$La$_{0.02}$TiO$_3$ resistivity (2 mΩ-cm)—STO needs only to be doped to an oxygen vacancy concentration of ~6x10$^{17}$ cm$^{-3}$, or only 0.002%; in addition, a concentration this low results in a thermopower of ~900 μV/K,[9] close to the value reported[9,11] for that of two-dimensional thermoelectrics on STO substrates.

The fact that even lightly reduced STO substrates were contributing to our transport measurements is evidenced both by the orders-of-magnitude disparity between measurements of films on STO and of films on other substrates, and on the convergence to bulk transport values for thicker SLTO films grown on STO (Fig. 2). At thicknesses of ~ 1000 nm the film plays a dominant role in carrying current through the system, as thermopowers in any parallel-resistor-type system such as our film-substrate sample can be added linearly as weighted by the conductances of each layer.

Transport measurements were performed on our thin-films of SLTO by evaporating 25 nm chromium / 100nm gold Ohmic contacts in four-point Van der Pauw geometry. Thermopowers $S$ were obtained by mounting the samples between two Peltier devices to provide a temperature gradient across the substrate and film. T-type thermocouples were contacted to the Cr/Au sample contacts, and temperature and voltage were recorded with the same wires and so at the same point.

At growth pressures higher than $10^{-7}$ Torr (Fig. 3a) the SLTO films showed a much higher resistivity, $10^5$ m$\Omega$-cm, than SLTO in bulk form, and a correspondingly higher thermopower, 750 $\mu$V/K. This difference is not yet well understood. At the lowest growth pressures of $10^{-7}$ Torr, thermopower is even higher at >1 mV/K, while the measured sheet resistance of the system is six orders of magnitude lower than in the high-pressure samples (~12 $\Omega$). This strongly suggests that two materials are contributing to the overall measurements. This behavior was not observed on other substrates where the values of thermopower measured on films of $10^{-7}$ growth pressure were always close to the bulk value of 200-300 $\mu$V/K.

Sheet resistance was measured from the backside of various substrates with the SLTO film remaining on the frontside (Fig. 3b) similarly to the frontside measurements, using sputtered Ti/Pd rather than Cr/Au to give Ohmic contact. This yielded a value identical to that measured from the frontside (~12 $\Omega$/□), implying that the substrate either contains short-circuit paths perpendicular to the plane of the film or acts entirely as a conductive shunt. Sheet resistance measurements on the backsides of other substrates yielded values at the measurement limit of our apparatus (>35 MΩ).

The ostensibly surprising result that SrTiO$_3$ is strongly reduced during short PLD growth times at low temperatures (10-60 minutes at 450°C) to electrically short the film-substrate system is less curious when one accounts for two things: first, the extremely low doping level (<0.01%) of oxygen vacancies necessary for large thermopower in SrTiO$_3$, and second, the relative size of the substrate with respect to the thin film grown on it. It is evident that the PLD growth process itself induces the oxygen vacancies in the STO substrate, as the optical darkening of the substrate[24] (with the film removed) occurred even with the growth of films of undoped SrTiO$_3$. It has been suggested previously that the growth of oxide thin-films may reduce STO substrates[25,26]. In addition, the large observed variation in STO substrate quality and in turn a variability in dislocation density from substrate to substrate may play a role in the oxygen reduction mechanism during film growth.

If it were assumed that all of the conduction was occurring in the 100-200 nm SLTO films grown at 10$^{-7}$ Torr, a resistivity of 1 mΩ-cm and thermopower of ~ 1 mV/K would yield a power factor at 300K of 30 W-m$^{-1}$K$^{-1}$, or >10 times that of Bismuth Telluride. It should be noted that this is not an unphysical result. The lattice distortion and EELS spectra (not shown) of the SLTO films indicate that they, too, are reduced to Sr$_{0.98}$La$_{0.02}$TiO$_{3-\delta}$, and therefore the introduction of an oxygen-vacancy band may play a role in transport. A parabolic-band Boltzmann transport model[27] can be used to approximate the thermopower and resistivity in Sr$_{1-x}$La$_x$TiO$_{3-\delta}$ films given appropriate inputs for scattering time $\tau$, scattering mechanism $r$, carrier concentration $n$, and carrier effective mass $m^*$. The two models one can consider are of the form $\tau(E) \sim E^{r-1/2}$, one with the exponent $r = 1/2$ for acoustic phonon scattering and the other with $r = 2$ for ionized impurity scattering, including scattering off oxygen vacancies. For the relaxation time $\tau = 45$ fs, previously reported[9] in SrTiO$_{3-\delta}$, we find that a power factor $S^2\sigma$ as high as 30 W-m$^{-1}$K$^{-1}$ is consistent only if the carrier concentration $n < 1 \times 10^{18}$ cm$^{-3}$, and the effective mass $m^* \approx 16m_e$ if ionized impurity scattering dominates ($r = 2$). It has been recently suggested[28] that ionized impurity scattering may already be present in Sr$_{1-x}$La$_x$TiO$_3$, and this value of $m^*=16m_e$ is consistent with the same previously reported data. Since the La-doped films will contain a minimum carrier concentration of $10^{20}$ cm$^{-3}$, a power factor of this magnitude would thus be an unphysical result without both an effective mass greater than $16m_e$ and a scattering time greater than 45fs. Conversely, a lightly-doped STO substrate with these values of $m^*$ and $\tau$ can achieve high thermopower (~1 mV/K) due to a lower carrier concentration from oxygen vacancies only.

The choice of substrate poses a dilemma for thin-film studies of doped STO systems. STO substrates, ideally suited for the growth of high-quality thin-films, appreciably reduce during thin-film growth to convolute thermoelectric measurements. However, the remarkably high thermopower that oxygen vacancy-doped STO can achieve at room temperature in bulk means that further refinement of the band structure, through careful selection of dopants and their concentrations, may offer a pathway to materials with $zT$ greater than one.

**Acknowledgments**

This work was supported by the U.S. Department of Energy. The authors also acknowledge support of the staff and facilities at the National Center for Electron Microscopy, Lawrence Berkeley National Laboratory, which is supported by the U.S. DOE under Contract No. DE-AC02-05CH112.

**Figure captions**

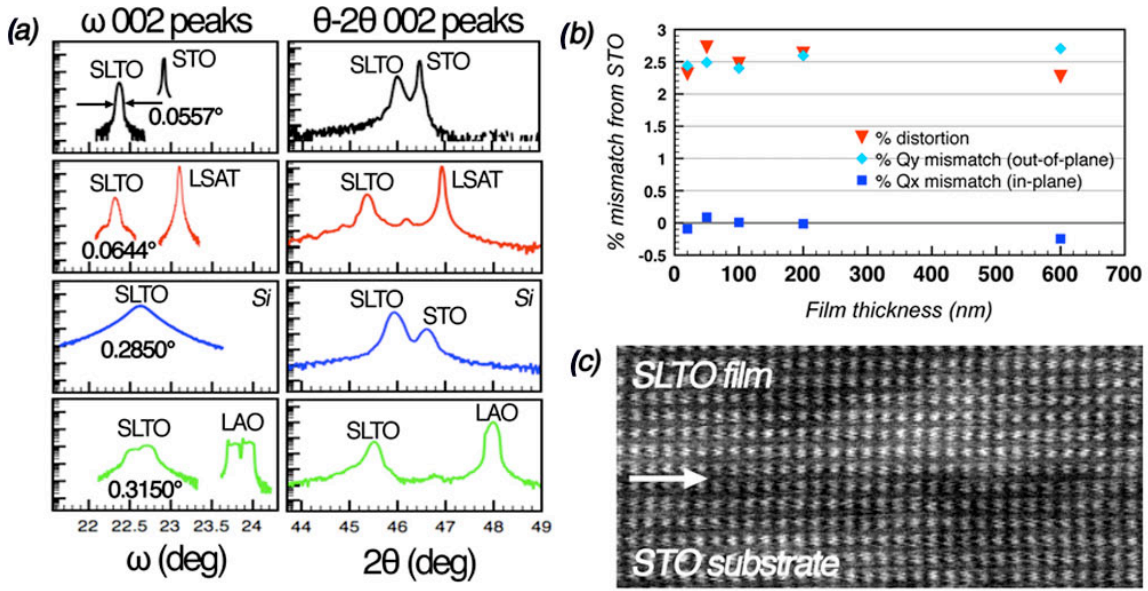

**FIG 1.** Structural data for $Sr_{0.98}La_{0.02}TiO_3$ (SLTO) (001) thin-films grown on STO (001), LSAT (001), STO (001)-on-Silicon (110) (Si), and LAO (001) substrates at $p_{O2} = 10^{-7}$ Torr and T = 450°C. A) Film and substrate rocking-curves and θ-2θ 002 diffraction peaks, with the full-width-at-half-maximum (FWHM) values of the films' 002 rocking-curve (ω) peaks noted. B) In-plane lattice parameters of SLTO and STO as determined from analysis of the 013 diffraction peaks. C) STEM image at the film/substrate interface.

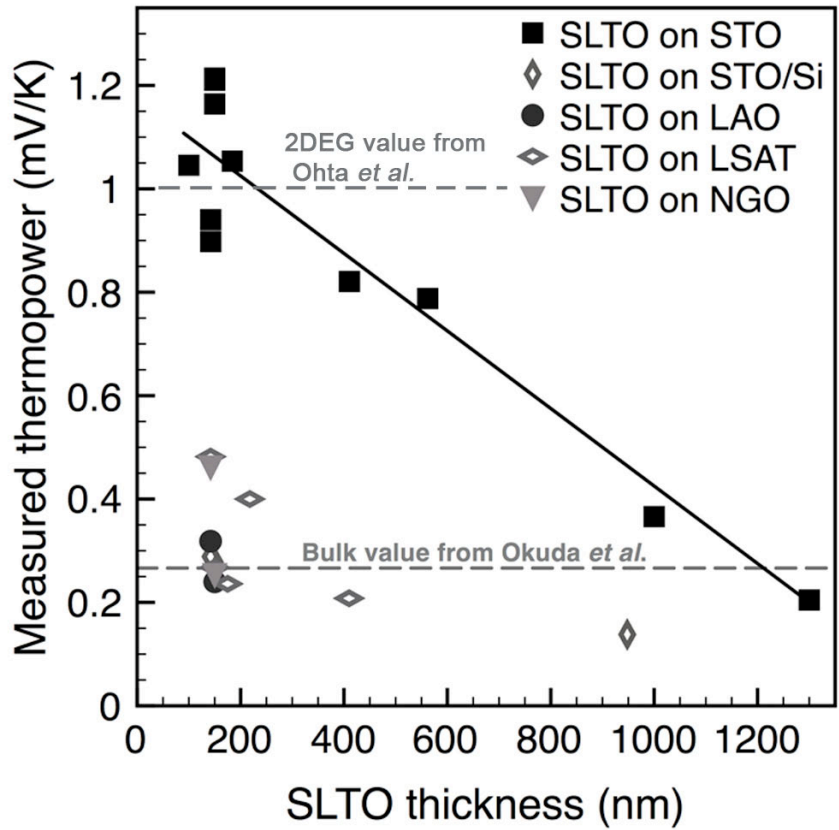

**FIG. 2**. Measured thermopower as a function of $Sr_{0.98}La_{0.02}TiO_3$ (SLTO) film thickness for films grown via pulsed laser deposition at 450°C and $p_{O2} = 10^{-7}$ Torr on STO, STO/Si, LAO, LSAT, and NGO substrates.

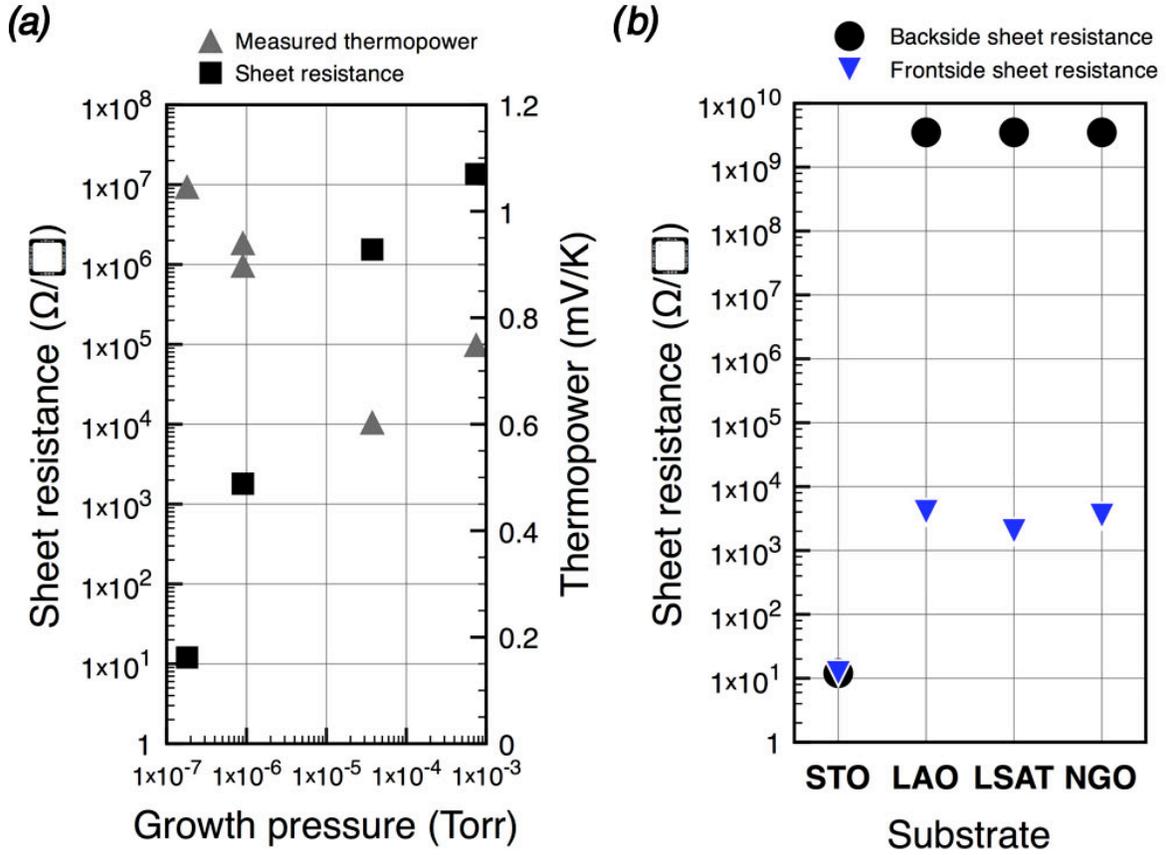

**FIG. 3.** Transport of 100 nm-thick $Sr_{0.98}La_{0.02}TiO_3$ thin-films as a function of growth pressure at 450°C on STO and as a function of substrate material at 450°C and $10^{-7}$ Torr growth pressure. A) Measured thermopower $S$ and measured sheet resistance $R_S$ as a function of growth pressure for SLTO films grown via PLD. B) Sheet resistances $R_S$ from the front- and backsides of our samples.